\documentclass[12pt,superscriptaddress]{revtex4}
\usepackage{graphicx}
\begin{document}
\begin{center}
{\bf\large Non-Gaussianity as a Probe of the Physics of the Primordial Universe and the Astrophysics of the Low Redshift Universe}
\vspace{2mm}
\thispagestyle{empty}

{\small
E. Komatsu,$^{1,2}$
N. Afshordi,$^3$
N. Bartolo,$^4$
D. Baumann,$^{5,6}$
J.R. Bond,$^7$
E.I. Buchbinder,$^3$
C.T.Byrnes,$^8$
X. Chen,$^9$
D.J.H. Chung,$^{10}$
A. Cooray,$^{11}$
P. Creminelli,$^{12}$
N. Dalal,$^7$
O. Dor\'e,$^7$
R. Easther,$^{13}$
A.V. Frolov,$^{14}$
K.M. G\'orski,$^{15}$
M.G. Jackson,$^{16}$
J. Khoury,$^{17}$
W.H. Kinney,$^{18}$
L. Kofman,$^7$
K. Koyama,$^{19}$
L. Leblond,$^{20}$
J.-L. Lehners,$^{21}$
J.E. Lidsey,$^{22}$
M. Liguori,$^{23}$
E.A. Lim,$^{24}$
A. Linde,$^{25}$
D.H. Lyth,$^{26}$
J. Maldacena,$^{27}$
S. Matarrese,$^{4,28}$
L. McAllister,$^{29}$
P. McDonald,$^7$
S. Mukohyama,$^2$
B. Ovrut,$^{17,27}$
H.V. Peiris,$^{30}$
C. R\"ath,$^{31}$
A. Riotto,$^{28,32}$
Y. Rodriguez,$^{33,34}$
M. Sasaki,$^{35}$
R. Scoccimarro,$^{36}$
D. Seery,$^{23}$
E. Sefusatti,$^{37}$
U. Seljak,$^{38,39}$
L. Senatore,$^{27}$
S. Shandera,$^{24}$
E.P.S. Shellard,$^{23}$
E. Silverstein,$^{25,40}$
A. Slosar,$^{41}$
K.M. Smith,$^{30}$
A.A. Starobinsky,$^{42}$
P.J. Steinhardt,$^{21,43}$
F. Takahashi,$^2$
M. Tegmark,$^{44}$
A.J. Tolley,$^3$
L. Verde,$^{45}$
B.D. Wandelt,$^{46}$
D. Wands,$^{19}$
S. Weinberg,$^{1,47}$
M. Wyman,$^3$
A.P.S. Yadav,$^5$
M. Zaldarriaga$^{5,6}$
}\\[0mm]
%{\footnotesize{\sf komatsu@astro.as.utexas.edu; 512-471-1483}}
\end{center}
\vspace{-3.5mm}
{\tiny$^1$ Texas Cosmology Center, The University of Texas at Austin,
Austin, TX 78712}\\ [-1.5mm]
{\tiny$^2$ IPMU, University of Tokyo, Chiba 277-8582, Japan}\\[-1.5mm]
{\tiny$^3$ Perimeter Institute for Theoretical Physics, Waterloo,
Ontario N2L 2Y5, Canada}\\[-1.5mm]
{\tiny$^4$ Dipartimento di Fisica, Universit\'a di, Padova, I-35131
Padua, Italy}\\[-1.5mm]
{\tiny$^5$ Center for Astrophysics, Harvard University, Cambridge, MA
02138}\\[-1.5mm]
{\tiny$^6$ Jefferson Physical Laboratory, Harvard University,
Cambridge, MA 02138}\\[-1.5mm]
{\tiny$^7$ CITA, University of Toronto, ON M5S 3H8, Canada}\\[-1.5mm]
{\tiny$^8$ Institut f\"ur Theoretische Physik, Universit\"at Heidelberg, 69120
Heidelberg, Germany}\\[-1.5mm]
{\tiny$^9$ Center for Theoretical Physics, MIT, Cambridge, MA 02139}\\[-1.5mm]
{\tiny$^{10}$ Department of Physics, University of Wisconsin, Madison, WI 53706}\\[-1.5mm]
{\tiny$^{11}$ Department of Physics and Astronomy, University of California, Irvine, CA 92697}\\[-1.5mm]
{\tiny$^{12}$ Abdus Salam International Center for Theoretical Physics, Trieste, Italy}\\[-1.5mm]
{\tiny$^{13}$ Department of Physics, Yale University, New Haven, CT
06520}\\[-1.5mm]
{\tiny$^{14}$ Department of Physics, Simon Fraser University, Burnaby,
BC, V5A 1S6, Canada}\\[-1.5mm]
{\tiny$^{15}$ JPL, Pasadena CA 91109; Caltech, Pasadena CA 91125; Warsaw
University Observatory, 00-478 Warszawa, Poland}\\[-1.5mm]
{\tiny$^{16}$ Lorentz Institute for Theoretical Physics, Leiden 2333CA,
The Netherlands}\\[-1.5mm] 
{\tiny$^{17}$ Department of Physics \& Astronomy, University of Pennsylvania, Philadelphia, PA 19104}\\[-1.5mm]
{\tiny$^{18}$ Department of Physics, University at Buffalo, SUNY, Buffalo, NY 14260}\\[-1.5mm]
{\tiny$^{19}$ Institute of Cosmology and Gravitation, University of Portsmouth, Portsmouth, UK}\\[-1.5mm]
{\tiny$^{20}$ Department of Physics, Texas A\&M University, College Station, TX 77843}\\[-1.5mm]
{\tiny$^{21}$ Princeton Center for Theoretical Sciences, Princeton University, Princeton, NJ 08544}\\[-1.5mm]
{\tiny$^{22}$ %Astronomy Unit, 
School of Mathematical Sciences, Queen Mary, University of London,  London E14NS, UK}\\[-1.5mm]
{\tiny$^{23}$ DAMTP, University of Cambridge, Cambridge CB3 0WA, UK}\\[-1.5mm]
{\tiny$^{24}$ ISCAP, Physics Department, Columbia University, New York, NY 10027}\\[-1.5mm]
{\tiny$^{25}$ Department of Physics, Stanford University, Stanford, CA 94305}\\[-1.5mm]
{\tiny$^{26}$ Department of Physics, Lancaster University, Lancaster LA1 4YB, UK}\\[-1.5mm]
{\tiny$^{27}$ School of Natural Sciences, Institute for Advanced Study, Princeton, NJ 08540}\\[-1.5mm]
{\tiny$^{28}$ INFN, Sezione di Padova, I-35131 Padua, Italy}\\[-1.5mm]
{\tiny$^{29}$ Department of Physics, Cornell University, Ithaca, NY 14853}\\[-1.5mm]
{\tiny$^{30}$ Institute of Astronomy, Cambridge University, Cambridge,
UK}\\[-1.5mm]
{\tiny$^{31}$ Max-Planck-Institut f\"ur Extraterrestrische Physik, 85748 Garching, Germany}\\[-1.5mm]
{\tiny$^{32}$ CERN, PH-TH Division, CH-1211, Geneve 23, Switzerland}\\[-1.5mm]
{\tiny$^{33}$ Centro de Investigaciones, Universidad Antonio
Nari\~no Cra 3 Este \# 47A-15, Bogot\'a D.C., Colombia}\\[-1.5mm]
{\tiny$^{34}$ Escuela de Fisica, Universidad Industrial de Santander Ciudad Universitaria, Bucaramanga, Colombia}\\[-1.5mm]
{\tiny$^{35}$ Yukawa Institute for Theoretical Physics,
Kyoto University, Kyoto 606-8502, Japan}\\[-1.5mm]
{\tiny$^{36}$ Center for Cosmology and Particle Physics,
Department of Physics, NYU, New York, NY 10003}\\[-1.5mm]
{\tiny$^{37}$ Institut de Physique Th\'eorique, CEA-Saclay, F-91191
Gif-sur-Yvette, France}\\[-1.5mm]
{\tiny$^{38}$ Physics and Astronomy Department, University
of California, and LBNL, Berkeley, CA 94720}\\[-1.5mm]
{\tiny$^{39}$ Institute for Theoretical Physics, University of Z\"urich, CH-8057
Z\"urich, Switzerland}\\[-1.5mm]
{\tiny$^{40}$ SLAC, Stanford University, Stanford, CA 94305}\\[-1.5mm]
{\tiny$^{41}$ Berkeley Center for Cosmological Physics, University
of California, Berkeley, CA 94720}\\[-1.5mm]
{\tiny$^{42}$ Landau Institute for Theoretical Physics, Moscow
119334, Russia}\\[-1.5mm] 
{\tiny$^{43}$ Joseph Henry Laboratories, Princeton University,
Princeton, NJ08544}\\[-1.5mm] 
{\tiny$^{44}$ Department of Physics and MIT Kavli Institute, MIT,
Cambridge, MA 02139}\\[-1.5mm]   
{\tiny$^{45}$ Institute of Space Sciences (IEEC-CSIC), Fac. Ciencies,
Campus UAB, Bellaterra, Spain}\\[-1.5mm] 
{\tiny$^{46}$ Departments of Physics and Astronomy,
University of Illinois at Urbana-Champaign, Urbana, IL 61801}\\[-1.5mm]  
{\tiny$^{47}$ Theory Group, Department of Physics, The University of Texas at Austin, Austin, TX 78712}
%%%%%%%%%%%%%%%%%%%%%%%%%%%%%%%%%%%%%%%%%%%%%%%%%%%%%%%%%%%%%%%%%%%%%
\newpage
\setcounter{page}{0}
\section*{EXECUTIVE SUMMARY}
A new and powerful probe of the origin and evolution of structures in
the Universe has emerged and been actively developed over the last
decade. In the coming decade, {\it
non-Gaussianity}, i.e., the study of non-Gaussian contributions to the
correlations of cosmological fluctuations, will become an important
probe of both the early and the late Universe. Specifically, it will
play a leading role in furthering our understanding of two fundamental
aspects of cosmology and astrophysics: 
\begin{itemize}
 \item  The physics of the very early universe that created the
	primordial seeds for large-scale structures, and
 \item  The subsequent growth of structures via gravitational
	instability and gas physics at later times.
\end{itemize}
To date, observations of fluctuations in the Cosmic Microwave Background
(CMB) and the Large-Scale Structure of the Universe (LSS) have focused
largely on the 
Gaussian contribution as measured by the two-point correlations (or the
power spectrum) of density fluctuations.  However, an even
greater amount of information is contained in non-Gaussianity and a
large discovery space therefore still remains to be explored. 
Many observational
probes can be used to measure non-Gaussianity, including CMB, LSS,
gravitational lensing, Lyman-$\alpha$ forest, 21-cm fluctuations, and
the abundance of 
rare objects such as clusters of galaxies 
and high-redshift galaxies. {\it Not only does the study of non-Gaussianity
maximize the science return from a plethora  of present and
future cosmological experiments and observations, but it also carries
great potential for important discoveries in the coming decade}.
%%%%%%%%%%%%%%%%%%%%%%%%%%%%%%%%%%%%%%%%%%%%%%%%%%%%%%%%%%%%%%%%%%%%%
\section{Beyond a simple approximation to nature}
The last decade has witnessed tremendous advances in our understanding of
the Universe. The measurements of the anisotropies in the CMB
temperature and polarization fluctuations by the Cosmic Background
Explorer (COBE), the Wilkinson Microwave Anisotropy Probe (WMAP), and
many ground-based and sub-orbital experiments, and of the distribution
of galaxies by the CfA Redshift Survey, Two-degree Field Galaxy Redshift
Survey (2dFGRS), and the Sloan Digital Sky Survey
(SDSS), among others, were milestones in modern cosmology.
The two-point correlation
function (or its Fourier transform, the power spectrum) of temperature
and polarization anisotropies, as well as that of the galaxy
distribution, have sharpened our view of the Universe
(e.g., \cite{komatsu/etal:2009}) - 
we now know that the Universe is $13.7\pm 0.1$~Gyr old, and made 
of $4.6\pm 0.2$\% hydrogen and helium nuclei, $22.8\pm 1.3$\%  dark
matter, and the rest, $72.6\pm 1.5$\%, is in the form of dark
energy. The spatial geometry of the observable
Friedmann-Robertson-Walker Universe is spatially flat (Euclidean) to
about 1\%. 

Explaining the CMB and LSS power spectra requires only a handful of
numbers: today's expansion rate, the energy density of atoms, dark matter 
and dark energy, the optical depth resulting from hydrogen reionization,
and the amplitude and 
scale-dependence of the primordial seed fluctuations. 
However, knowing the values of these parameters does not provide us with
a complete understanding of the physical laws governing the Universe: 
knowing the abundance of dark matter and dark 
energy does not tell us what they are. Knowing the amplitude and
scale-dependence of the primordial fluctuations does not tell us
what created those primordial fluctuations. It is  clear
that we need more information.

In fact, we {\it do} have more information: the WMAP temperature map contains
$10^6$ pixels, and there are $10^5$ spectra of galaxies surveyed by
SDSS. Yet, cosmologists spent the last decade measuring and interpreting the
two-point correlations, which contain only $\sim 1000$ and $100$ numbers
for WMAP and SDSS, respectively. 

This kind of compression of the data is justified if, and {\it only if}, the
statistical distribution of the observed fluctuations is a Gaussian
distribution with random phases. Any information contained in the
departure from a perfect Gaussian, {\it non-Gaussianity}, is not encoded
in the power 
spectrum, but has to be extracted from measurements of higher-order
correlation functions. 
The study and characterization of non-Gaussianity began three decades
ago with the first large scale structure surveys, but the main focus
until recently has been on two-point correlations and Gaussian fluctuations.

In this White Paper we describe how non-Gaussianity is a particularly
potent probe of the fundamental origin and the late time evolution of
structures. 

%%%%%%%%%%%%%%%%%%%%%%%%%%%%%%%%%%%%%%%%%%%%%%%%%%%%%%%%%%%%%%%%%%%%%
\section{Non-Gaussianity as a probe of the physics of the primordial Universe}
Over the last decade we have accumulated a good deal of observational
evidence from CMB and LSS power spectra that the observed
structures originated from seed fluctuations in the very early universe.
The leading theory explaining the primordial origin of cosmological
fluctuations is cosmic inflation \cite{guth:1981}, a period of
accelerated expansion at very early times. During inflation, 
microscopic quantum fluctuations were stretched to macroscopic scales to
provide the seed fluctuations for the formation of large-scale
structures like our own Galaxy.  

What was the physics responsible for inflation?
Many theoretical ideas have been proposed to explain 
the existence of an early phase of accelerated expansion. 
Inflation models with the minimum number of degrees of freedom,
parameters and tuning needed to solve the flatness and homogeneity
problems give a fairly well-defined range of predictions.
While the current experimental data has ruled out a good fraction of
that range, there remains a substantial range that still fits the
data \cite{komatsu/etal:2009}. 
	
Learning about the physics of inflation is equivalent to
learning about the evolution and interactions of quantum fields in the
very early Universe. Measurements of the power spectrum alone have limited
potential in revealing this information. The power spectrum is
determined by the inflationary expansion rate and its time-dependence
which in turn relates to the evolution of the inflationary energy
density. However, the power spectrum does not strongly constrain the
interactions of the field (or fields) associated with this energy
density.  The power spectrum is therefore degenerate in terms of the
inflationary action that can lead to it - 
the power spectrum is therefore degenerate in terms of the
inflationary action that can lead to it - inflation models with different field 
interactions can lead to very similar predictions for the power spectrum.
Non-Gaussianity is a sensitive probe of the aspects of inflation
that are difficult to probe by other means.
Specifically, it is a probe
of the interactions of the field(s) driving inflation and therefore
contains vital information about the fundamental physics operative
during inflation. 

In many single field slow-roll models the non-Gaussianity is small and likely
unobservable by virtue of the inflaton field being weakly
coupled. However, a large, detectable amount of non-Gaussianity can be
produced when any of the following conditions is violated:
\begin{itemize}
 \item {\bf Single Field}. There was only one quantum field responsible
       for driving inflation and for generating the primordial seeds for
       structures. 
 \item {\bf Canonical Kinetic Energy}. The kinetic energy of the quantum
       field is such that the speed of propagation of fluctuations is
       equal to the speed of light. 
 \item {\bf Slow Roll}. The evolution of the field was always very slow
       compared to the Hubble time during inflation.
 \item {\bf Initial Vacuum State}. 
       %Spatial Fourier modes of the quantum 
       %field which are observable now were in the preferred adiabatic
       %vacuum state (also sometimes called the ``Bunch-Davies vacuum'') just
       %before the quantum fluctuations were generated during
       %inflation. 
       The quantum field was in the preferred adiabatic
       vacuum state (also sometimes called the ``Bunch-Davies vacuum'') just
       before the quantum fluctuations were generated during inflation. 
\end{itemize}
Inflation is expected to produce undetectable levels of primordial
non-Gaussianity, only when {\it all} of the above conditions are
satisfied (see, e.g., \cite{bartolo/etal:2004}, for a review) - the
conditions that inflation models have the minimum number of degrees of
freedom, parameters and tuning needed to solve the flatness and
homogeneity problem. {\it Confirming or ruling out this class of inflation
models is an important goal}.

\begin{figure}
\centering \noindent
\includegraphics[width=16cm]{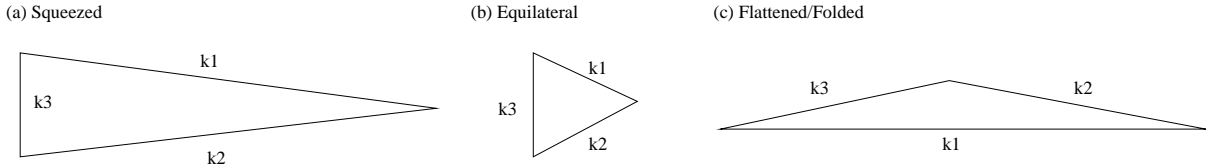}
\caption{%
Bispectrum shapes, 
 $B(k_1,k_2,k_3)$, which can be characterized by triangles formed by three
 wave vectors. The shape (a) has the maximum signal at the squeezed
 configuration, $k_3\ll k_2\approx k_1$, and can be produced by
 models of inflation involving multiple fields. The shape (b) has the
 maximum signal 
 at the equilateral configuration, $k_1=k_2=k_3$, and can be produced
 by non-canonical kinetic terms of quantum fields. The shape (c) has the
 maximum signal at the flattened configuration, $k_1\approx 2k_2\approx 2k_3$, and can be produced by non-vacuum initial conditions.
}
\label{fig:triangles}
\end{figure} 

Non-Gaussianity is measured by various methods. A standard approach is
to measure non-Gaussian correlations, i.e., the correlations that vanish
for a Gaussian distribution. The three-point function (or its
Fourier transform, the bispectrum) is such a correlation.  

The three-point function correlates density or temperature fluctuations
at three points in space. Equivalently, the bispectrum,
$B(k_1,k_2,k_3)$, correlates 
fluctuations with three wave vectors (see Figure~\ref{fig:triangles}).
These three wave vectors form a
triangle in Fourier space, and thus there are many triangles one can
form and look for. The amount of information captured by the
bispectrum is therefore potentially far greater than that of the power
spectrum, which correlates only two wave vectors with the same magnitude.

An important theoretical discovery
made toward the end of the last decade is that violation of each of the
above conditions (single field, canonical kinetic energy, slow roll,
and initial vacuum state) 
results in unique signals with specific triangular shapes: multi-field
models, non-canonical kinetic term models, 
non-adiabatic-vacuum models
(e.g., initially excited states), and non-slow-roll models can generate
signals in squeezed triangles ($k_3\ll k_2\approx k_1$),
equilateral triangles ($k_1=k_2=k_3$), flattened/folded triangles
($k_3\approx k_2\approx 2k_1$), and more complex configurations,
respectively (see, e.g.,
\cite{linde/mukhanov:1997,lyth/ungarelli/wands:2003,babich/creminelli/zaldarriaga:2004,chen/etal:2007,chen/easther/lim:2007,holman/tolley:2008}).
When more than one of the conditions are violated, a linear combination of
different shapes would 
arise \cite{langlois/etal:2008}.

The squeezed configuration in Fourier space is equivalent to 
the primordial curvature perturbation in position space, $\Phi({\mathbf
x})$  (up to a sign this is the usual Newtonian potential), 
given by $\Phi({\mathbf x})=\phi_g({\mathbf 
x})+f_{NL}\phi_g^2({\mathbf x})$, where $\phi_g({\mathbf x})$ is a
Gaussian field. This form of non-linearity was first recognized by
Ref. \cite{salopek/bond:1990} within the context of inflation, and the
parameter $f_{NL}$ characterizes the amount of non-Gaussianity in this
particular configuration. 
The latest constraint on
$f_{NL}$ from the WMAP 5-year
data is $f_{NL}=38\pm 21$ (68\%~CL; \cite{smith/etal:2009}). While the
statistical significance of the signal is still low (about 2-$\sigma$
level), future experiments such as the Planck CMB satellite and
high-redshift galaxy surveys are expected to yield much tighter constraints
\cite{komatsu/spergel:2001,sefusatti/komatsu:2007}, and might well lead
to a convincing detection. 

A new  method for
measuring $f_{NL}$ from galaxy surveys that does not rely on the bispectrum, but
uses the fact that the power spectrum of density extrema (where galaxies
are formed) on large scales increases (decreases) for a positive
(negative) $f_{NL}$ \cite{dalal/etal:2008} (also see
\cite{matarrese/verde:2008} for a generalized result) is particularly 
promising. More specifically, $f_{NL}$ introduces a scale-dependent
modification of the galaxy power spectrum, which increases as $\sim
1/k^2$ as one goes to smaller $k$ (larger spatial scales), and
evolves roughly as $(1+z)$ as a function of redshift.
This method yields a competitive limit already from SDSS
\cite{slosar/etal:2008}, and there is a realistic chance that one can
reach sensitivity at the level of $\Delta f_{NL}\lesssim
1$, e.g., \cite{seljak:2009,mcdonald/seljak:2008,carbone/etal:2008}. Note that the signature
of non-Gaussianity is a smooth feature; thus, wide-field photometric
surveys are well suited to study this effect.

These findings suggest that non-Gaussian correlations are a very
powerful probe of the physics of inflation. 
Understanding non-Gaussianity does
for inflation what direct detection experiments do for dark matter, or
the Large Hadron Collider for the Higgs particles. It probes the
interactions of the field sourcing inflation, revealing the
fundamental aspects of the physics at very high energies that are
not accessible to any collider experiments. For this reason, non-Gaussianity
has been a key player in the recent surge in a very productive exchange
of ideas between cosmologists and high-energy theorists, and we have
every reason to expect that this will continue in a bigger form in the
coming decade.  

Moreover, recent studies suggest
that potential alternatives to inflation scenarios, such as 
an early contracting phase of the Universe
followed by a bounce (rather than expanding), tend to
generate large non-Gaussianity. 
Null detection of non-Gaussianity at the level of $\Delta f_{NL}\lesssim
1$ would rule out all of the alternative models based on a contracting
phase currently proposed and reviewed in \cite{lehners:2008}. 

While detection of large non-Gaussianity would not rule out inflation,
it would rule out the class 
of models satisfying all of the above conditions simultaneously (single field,
canonical kinetic energy, slow roll, and initial vacuum state).
The most important aspect of
primordial non-Gaussianity is that a convincing 
detection of the
squeezed configuration, $f_{NL}$, will {\it rule out all classes of
inflationary models based upon a single field}
\cite{creminelli/zaldarriaga:2004}. The shape  of the two-point correlation
function (characterized by the so-called primordial tilt,
$n_s$, and the running index, $\alpha_s$) and the existence or absence
of primordial gravitational waves, would provide important constraints
on large classes of 
inflationary models, but they would never be able to rule out
single-field inflation. 

To summarize, non-Gaussian correlations offer a new window into the
details of the fundamental physics  of the
primordial Universe that are not accessible by Gaussian correlations.
%%%%%%%%%%%%%%%%%%%%%%%%%%%%%%%%%%%%%%%%%%%%%%%%%%%%%%%%%%%%%%%%%%%%%
\section{Non-Gaussianity as a probe of the astrophysics of the low-redshift Universe}
Gaussian fluctuations become non-Gaussian as cosmic structures
evolve and go through various non-linear processes.
{\it This property makes non-Gaussianity a sensitive
probe of the evolution of cosmic structures and numerous non-linear
astrophysical processes of the low-redshift Universe}.  

Non-Gaussianity
can be used to extract additional
information about the gravitational lensing effect
\cite{zaldarriaga/seljak:1999}, the Sunyaev--Zel'dovich effect
\cite{cooray:2000},  the cosmic reionization epoch
\cite{cooray/hu:2000}, and the Integrated Sachs--Wolfe effect
\cite{munshi/etal:1995,goldberg/spergel:1999}, which can be used to
constrain the 
equation of state parameter of dark energy \cite{verde/spergel:2000}.

Non-Gaussianity is a sensitive probe of small 
non-linear effects that must have existed at the photon decoupling
epoch, $z\simeq 1090$, via non-linear general relativistic effects
\cite{bartolo/matarrese/riotto:2004}, the non-linear evolution of
the photon-baryon fluid \cite{bartolo/matarrese/riotto:2006,pitrou:2009},
non-linear perturbations of the electron density at recombination
\cite{khatri/wandelt:2009,senatore/tassev/zaldarriaga:2008}, and
non-linearities in the 
radiative transfer such as the non-linear Sachs--Wolfe effect and weak lensing
\cite{pyne/carroll:1996}. 

Non-Gaussianity also offers powerful diagnostics of galaxy formation
measuring how galaxies trace the underlying mass
distribution (see \cite{bernardeau/etal:2002} for a review), as well as
of the physics of 
the Inter Galactic Medium (IGM) measuring how gas
traces the underlying mass distribution \cite{viel/etal:2004}. 
New tracers of the underlying mass distribution, the cosmological 21-cm
fluctuations (see \cite{furlanetto/oh/briggs:2006} for a review), will
contain far more information in its higher order 
correlation functions than in the two-point correlations. Further
theoretical studies are needed to exploit the rich information available in
the 21-cm fluctuations.

%%%%%%%%%%%%%%%%%%%%%%%%%%%%%%%%%%%%%%%%%%%%%%%%%%%%%%%%%%%%%%%%%%%%%
\section{How to exploit non-Gaussianity in the coming decade}
The tremendous power of non-Gaussianity for constraining the physics of
the primordial Universe and the astrophysics of the low redshift Universe
has begun to be fully appreciated toward the end of the last
decade. What do we expect over the next decade?

The theoretical discovery that different triangle configurations of the
bispectrum are sensitive to different aspects of the physics of
inflation was a major achievement of the last decade. 
So far three distinct configurations
(see Figure~\ref{fig:triangles}) have been investigated, but it is
entirely possible that new physics may be probed by different
configurations. Moreover, there is no reason to stop at the three-point
function. Recent studies suggest that the four-point function (or its
Fourier transform, the trispectrum) gives us additional information
about inflation models \cite{huang/shiu:2006} and
potential alternatives \cite{buchbinder/etal:2008}, beyond what is possible with
the three-point function, and the Planck CMB satellite is expected to
yield useful limits \cite{kogo/komatsu:2006}. Studies of
what is possible beyond the three-point function have just begun.
More theoretical studies are necessary to fully exploit the potential of
non-Gaussianity. 

Low redshift non-linear astrophysical phenomena are very rich
and important subjects by themselves; however, they may mask the primordial
non-Gaussian signatures. While several studies have suggested that the
squeezed configuration, $f_{NL}$, is relatively insensitive to low
redshift phenomena, e.g., \cite{serra/cooray:2008}, more studies are
required to develop a secure method to extract the primordial
non-Gaussianity. The low redshift contamination of the other triangle
configurations, as well as to the four-point function, is yet to be studied. 

The Galactic emission is non-Gaussian, and its effect
must be understood and subtracted. Studies of the
WMAP data \cite{komatsu/etal:2009} have shown that the Galactic
contamination of $f_{NL}$ is not very large; however, at the level of
sensitivity that the Planck satellite is expected to reach, $\Delta
f_{NL}\sim 5$, foregrounds  would play an important
role. Again, the foreground contamination of the other configurations
and the four-point function is yet to be studied. The method based upon
the galaxy power spectrum \cite{dalal/etal:2008} is still quite new, and
we need more investigations of the systematic errors in
this method to fully explot its potential of reaching $\Delta
f_{NL}\lesssim 1$. 

What kind of observations are needed for measurements of
non-Gaussianity? 
A sensible approach seems to measure non-Gaussianity with a combination
of many complementary observables including CMB, 
LSS, gravitational 
lensing, Lyman-$\alpha$ forest, 21-cm fluctuations, and the abundance of
clusters of galaxies and high-redshift
galaxies. Examples of on-going/funded missions include the Planck
satellite (CMB), the South Pole Telescope (SPT; CMB, clusters),
the Atacama Cosmology Telescope (ACT; CMB, clusters), the SDSS-III
(LSS, Ly$\alpha$ forest), the Hobby-Eberly Telescope Dark Energy Experiment (HETDEX;
LSS), the Dark Energy 
Survey (DES; LSS, clusters), the Panoramic Survey Telescope \& Rapid
Response System (Pan-STARRS; LSS, lensing), 
and the Extended R\"ontgen Survey with an
Imaging Telescope 
Array (eROSITA; LSS, clusters). Proposed future missions
include the Joint Dark Energy Mission (JDEM; LSS, lensing),
the Square Kilometer Array (SKA; LSS, 21-cm), the Large Synoptic Survey
Telescope (LSST; LSS, lensing), the Cosmic Inflation Probe (CIP; LSS), and a
CMB Polarization Satellite (CMBPol; CMB).

The variety of observations listed above are complementary in very
important ways: they probe different spatial scales (CMB probes the largest
spatial scales, whereas LSS, Ly$\alpha$, lensing, 21-cm, and clusters
probe small spatial scales that are not accessible with CMB) and fold
in a variety of post-inflationary physics. Such a 
range of observations expands the window opened by the CMB and may
uncover an unexpected interplay between cosmological phenomena in the
dynamics of evolving structures. 

It is overwhelmingly clear that theoretical advances in our
understanding of sources of observable non-Gaussianity will enable us to
extract much more information, and maximize the science return 
from a plethora of experiments.
%%%%%%%%%%%%%%%%%%%%%%%%%%%%%%%%%%%%%%%%%%%%%%%%%%%%%%%%%%%%%%%%%%%%%
\section{Conclusion}
Non-Gaussianity offers a powerful probe of the physics of the primordial
Universe and the non-linear
astrophysical processes in the low redshift Universe. Over the last
decade we have come to
realize the tremendous discovery potential of non-Gaussianity.
{\it Just about every on-going/funded/proposed cosmological
observation can be used effectively to measure non-Gaussianity, and
possibly revolutionize our understanding of the Universe in the past and
present}. 
%%%%%%%%%%%%%%%%%%%%%%%%%%%%%%%%%%%%%%%%%%%%%%%%%%%%%%%%%%%%%%%%%%%%%

\end{document}